\documentclass[conference]{IEEEtran}
\IEEEoverridecommandlockouts
\usepackage{cite}
\usepackage{amsmath,amssymb,amsfonts}
\usepackage{graphicx}
\usepackage{textcomp}
\usepackage{xcolor}
\usepackage{algorithm}
\usepackage{algorithmic}
\def\BibTeX{{\rm B\kern-.05em{\sc i\kern-.025em b}\kern-.08em
    T\kern-.1667em\lower.7ex\hbox{E}\kern-.125emX}}

\begin{document}

\title{OpenMENA: An Open-Source Memristor Interfacing and Compute Board for Neuromorphic Edge-AI Applications \\

\thanks{A. Safa supervised the project as Principal Investigator, designed the OpenMENA system, contributed to the experiments and writing of the manuscript. F. Mohsen contributed to the writing of the manuscript. Z. Ali contributed to the experiments and the writing of the manuscript, B. Wang and A. Bermak provided technical feedback and contributed to the writing of the manuscript. }
}

\author{\IEEEauthorblockN{Ali Safa, Farida Mohsen, Zainab Ali, Bo Wang, Amine Bermak}
\IEEEauthorblockA{$^1$\textit{College of Science and Engineering, Hamad Bin Khalifa University, Doha, Qatar} \\
asafa@hbku.edu.qa
}
}

\maketitle

\begin{abstract}
Memristive crossbars enable in-memory multiply–accumulate and local plasticity learning, offering a path to energy-efficient edge AI.  To this end, we present OpenMENA (Open Mimristor-in-Memory Accelerator), which, to our knowledge, is the first fully open memristor interfacing system integrating (i) a reproducible hardware interface for memristor crossbars with mixed-signal read–program–verify loops; (ii) a firmware–software stack with high-level APIs for inference and on-device learning; and (iii) a Voltage-Incremental Proportional–Integral (VIPI) method to program pre-trained weights into analog conductances, followed by chip-in-the-loop fine-tuning to mitigate device non-idealities. OpenMENA is validated on digit recognition, demonstrating the flow from weight transfer to on-device adaptation, and on a real-world robot obstacle-avoidance task, where the memristor-based model learns to map localization inputs to motor commands. OpenMENA is released as open source to democratize memristor-enabled edge-AI research.

\end{abstract}



\section*{Supplementary Material}

We release all hardware design and software material as open source at: \texttt{https://tinyurl.com/mr592wuf}

\section{Introduction}

Memristive devices are recognized as essential components for neuromorphic computing due to their non-volatility, analog conductance, and compact crossbar connectivity that enables in-memory multiply–accumulate (MAC) operations \cite{Sebastian2020NatNano}. Following the 2008 nanoscale realization of the “missing” memristor \cite{strukov2008missing}, resistive-switching memories (RRAM) and related variants have matured into compute-in-memory substrates that reduce data movement and improve performance-per-watt relative to von Neumann processors \cite{wan2022compute,Sebastian2020NatNano}. Recent work and system-level prototypes underscore potential of such implementations while highlighting the practical challenges mainly related to endurance limits, device variability, limited analog linearity, and peripheral circuit overheads \cite{Sebastian2020NatNano,ielmini2018memory,wan2022compute}. Recent  contributions analyze edge-oriented design opportunities and constraints for resistive RAM (ReRAM)-based compute-in-memory (CIM) architectures \cite{Singh2021ISCAS,McDanel2021ISCAS}.

Beyond efficient inference, memristors natively support \emph{local} synaptic plasticity. Spike-timing–dependent plasticity (STDP) and Hebbian learning \cite{9892362} can be implemented directly via pairs of pre- and post-synaptic spikes that induce conductance potentiation or depression \cite{BiPoo1998JNeurosci,jo2010nanoscale}, enabling unsupervised feature learning and online adaptation without the global error-backpropagation data traffic that dominates energy and area consumption in conventional training pipelines \cite{Sze2017ProcIEEE}. Device- and system-level studies have reproduced STDP in metal-oxide memristors \cite{jo2010nanoscale} and demonstrated \emph{in situ} learning on integrated crossbars \cite{prezioso2015training}, with recent hardware reports broadening the algorithm–hardware co-design space for robust learning under analog non-idealities \cite{aguirre2024hardware, 10548355}.
\begin{figure}[t]
    \centering
    \includegraphics[scale=0.32]{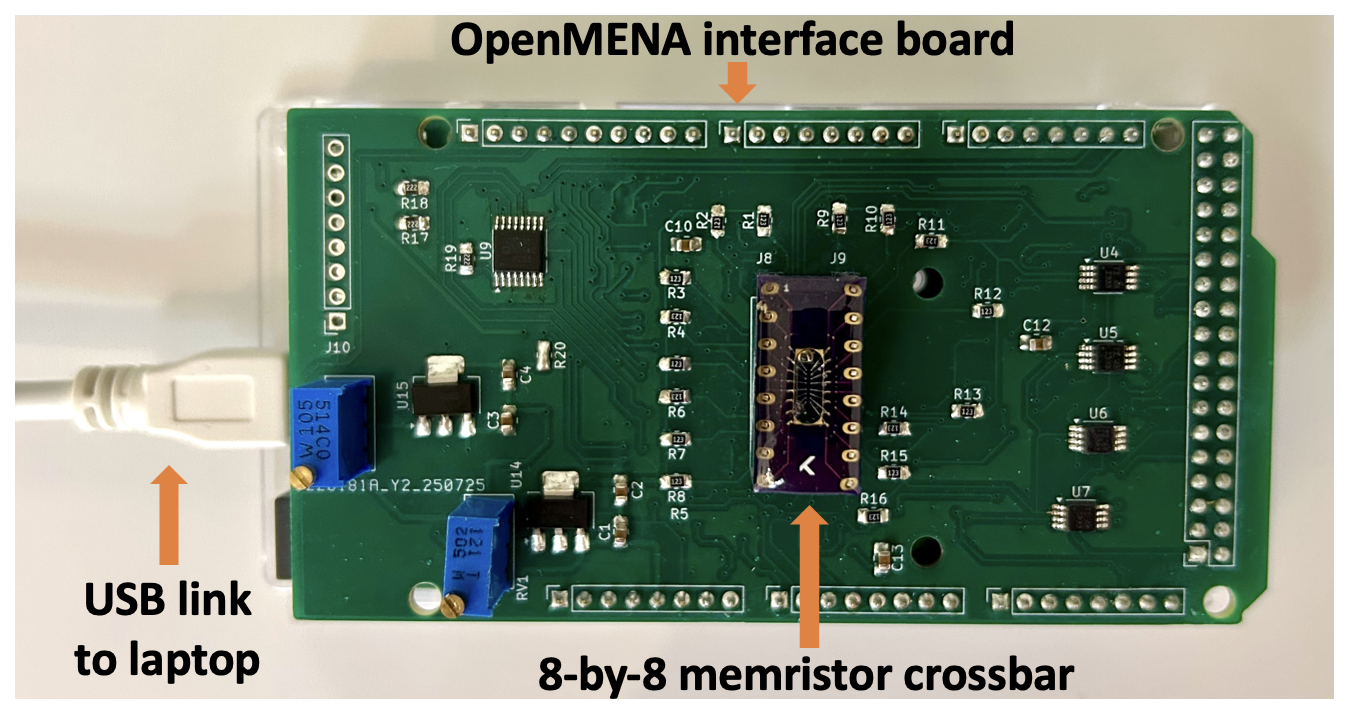}
    \caption{\textit{\textbf{The proposed OpenMENA system for memristor crossbar control and interfacing.} The OpenMENA PCB is mounted on an Arduino Due board for general purpose digital control. An 8-by-8 knowm memristor crossbar is mounted on the OpenMENA socket. The complete system is interface via a companion python library featuring both inference and weight setting functionalities. All code and design files are released as open-source to help democratize research in memristor-based neuromorphic AI.}}
    \label{fig:OpenMENA}
\end{figure}


However, advancing the field requires open, reproducible hardware–software platforms that allow researchers to span devices, circuits, learning rules, and applications under realistic constraints \cite{10771590}. Several useful systems exist, but they fall short of providing a \textit{fully open-source} stack targeted at embedded neuromorphic computing. For example, the knowm “Memristor Discovery” ecosystem offers open-source control software for benchtop characterization and education, yet the closed-source hardware modules and form factor prioritize desktop instrumentation rather than integrated edge processing \cite{KnowmGitHubDiscovery,KnowmProductPage}. Other efforts report impressive compute-in-memory chips and crossbar demonstrations but remain proprietary or release only partial artifacts \cite{wan2022compute,Sebastian2020NatNano}. Consequently, fully open-source, edge-oriented memristor development boards that integrate device interfacing, on-board compute, and learning-capable software remain scarce \cite{KnowmGitHubDiscovery,KnowmProductPage,wan2022compute,Sebastian2020NatNano}. In addition, prior work have documented practical challenges in crossbar programming effort and device endurance, reinforcing the need for open infrastructures that enable rapid and reproducible experimentation \cite{Farias2025ISCAS,McDanel2021ISCAS}.

To this end, this paper introduces OpenMENA (Open-source Memristor Edge-based Neuromorphic Architecture), which is, to our knowledge, the first open-source and reproducible platform available to the community. The contributions of this paper are as follows:

\begin{enumerate}
    \item We design and release a fully open-source system for interfacing memristor crossbars, which can be used for both AI inference and training of the memristor array.

    \item We introduce inference and training algorithms as part of OpenMENA's software suite using a proposed Voltage-Incremental Proportional-Integral (VIPI) control scheme for programming pre-trained weights into the memristor crossbar, followed by chip-in-the-loop fine-tuning.
    
    \item We validate OpenMENA end-to-end on both a digit-recognition benchmark and on a real-world robot obstacle-avoidance task. 

\end{enumerate}

This paper is organized as follows. OpenMENA's system design is described in Section \ref{systemdesign}. The various training and inference algorithms introduced in this work are presented in Section \ref{traininf}. Experimental results are shown in Section \ref{expres}. Finally, conclusions are provided in Section \ref{conc}.

\section{System Design}
\label{systemdesign}
\begin{figure}[htbp]
    \centering
    \includegraphics[scale=0.38]{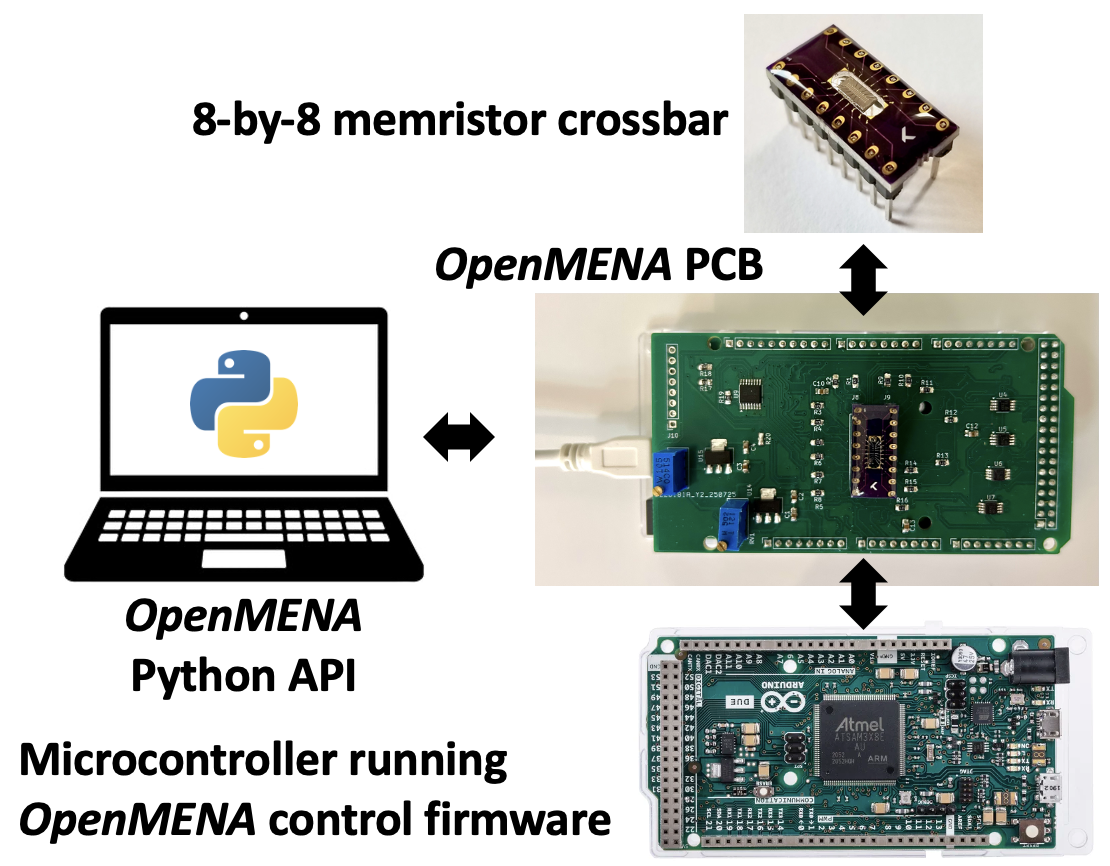}
    \caption{\textit{\textbf{Block diagram of the proposed OpenMENA system.} }}
    \label{fig:blockdia}
\end{figure}

The proposed OpenMENA system consists of \textit{i)} a memristor crossbar interfacing PCB that is mounted on top of an Arduino Due board and features a socket where crossbar modules can be plugged; \textit{ii)} the Arduino Due board runs the OpenMENA control firmware and the analog switches (used for both inference and weight setting), the digital-to-analog (DAC), and the output readouts via the Arduino Due's ADC channels; and \textit{iii)} the OpenMENA \textit{python} API that enables programming and control via python scripts running in an external laptop (connected via USB to the OpenMENA hardware). Fig. \ref{fig:blockdia} shows the block diagram of the proposed OpenMENA system.

\subsection{Memristor control and interfacing circuit}

\begin{figure}[htbp]
    \centering
    \includegraphics[scale=0.3]{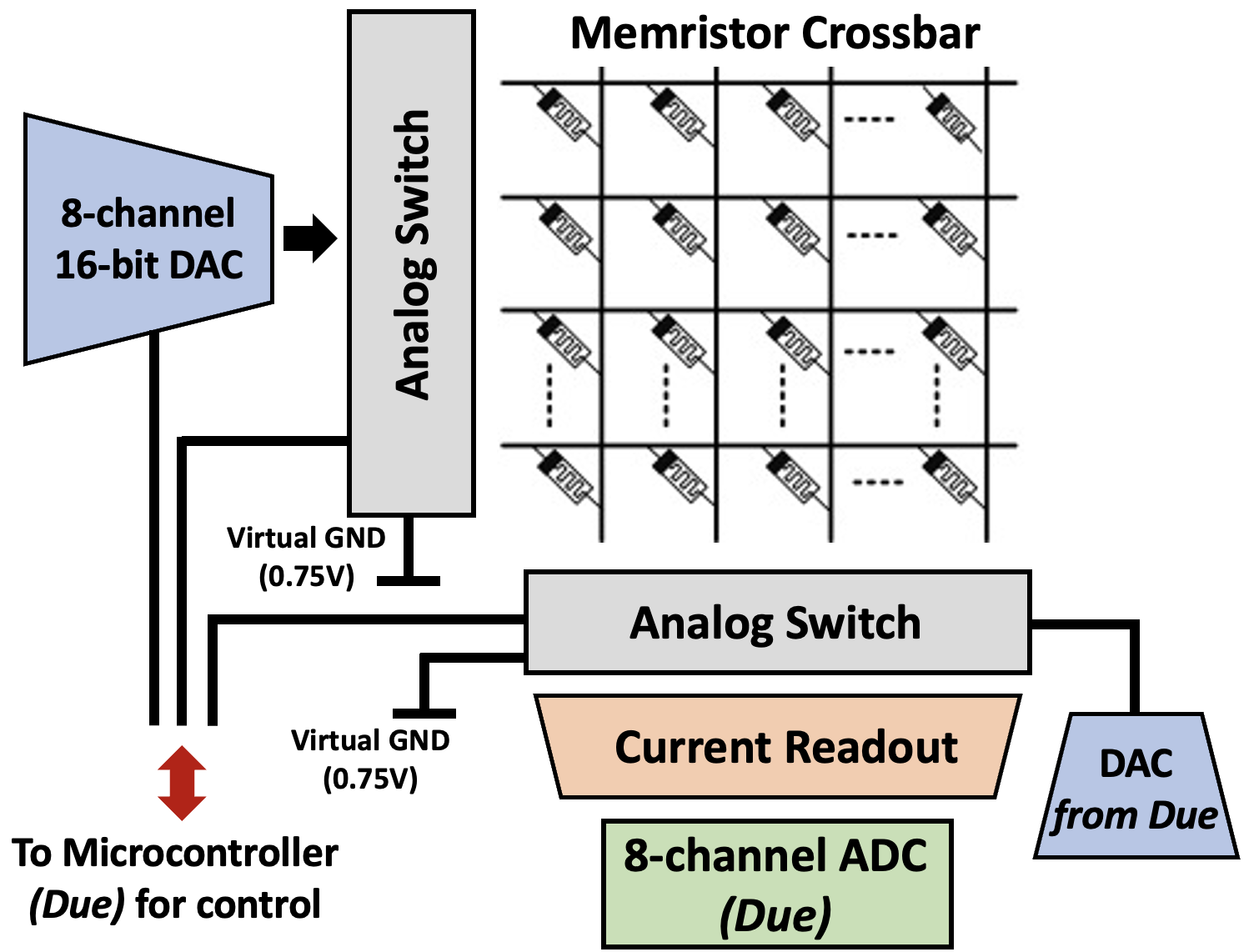}
    \caption{\textit{\textbf{Memristor crossbar control and readout circuit.} }}
    \label{fig:circuit}
\end{figure}

The OpenMENA PCB implements the memristor interfacing circuit shown in Fig. \ref{fig:circuit}  of an 8-channel DAC feeding into the memristor crossbar input via analog switches. The analog switches enable the writing of conductance values (acting as neural network \textit{weights}) into each memristor, by controlling voltage pulse signals modulating the memristor conductance values. In \textit{inference mode}, upon the application of input voltages (kept below the typical memristor switching threshold $V_{th}\approx 0.05\,\mathrm{V}$, with device resistances in the $\sim 10\,\mathrm{k}\Omega$–$1\,\mathrm{M}\Omega$ range), the resulting current at the output of the crossbar is routed to the current readout circuit (consisting of shunt resistors and instrumentation amplifiers) via a second set of analog switches.
 Finally, the voltage outputs of the current readout circuit are fed into the ADC channels of the microcontroller board for reading out the inference result. Positive and negative polarities are provided to the memristor crossbar by biasing the virtual ground level of the circuit to $0.75$V and providing voltage swings between $0$ and $1.5$V via the DACs. A complete schematic of the OpenMENA board is provided in the Supplementary Materials.


\subsection{Memristor conductance control and readout signals}
\label{bipolartech}
\begin{figure}[htbp]
    \centering
    \includegraphics[scale=0.37]{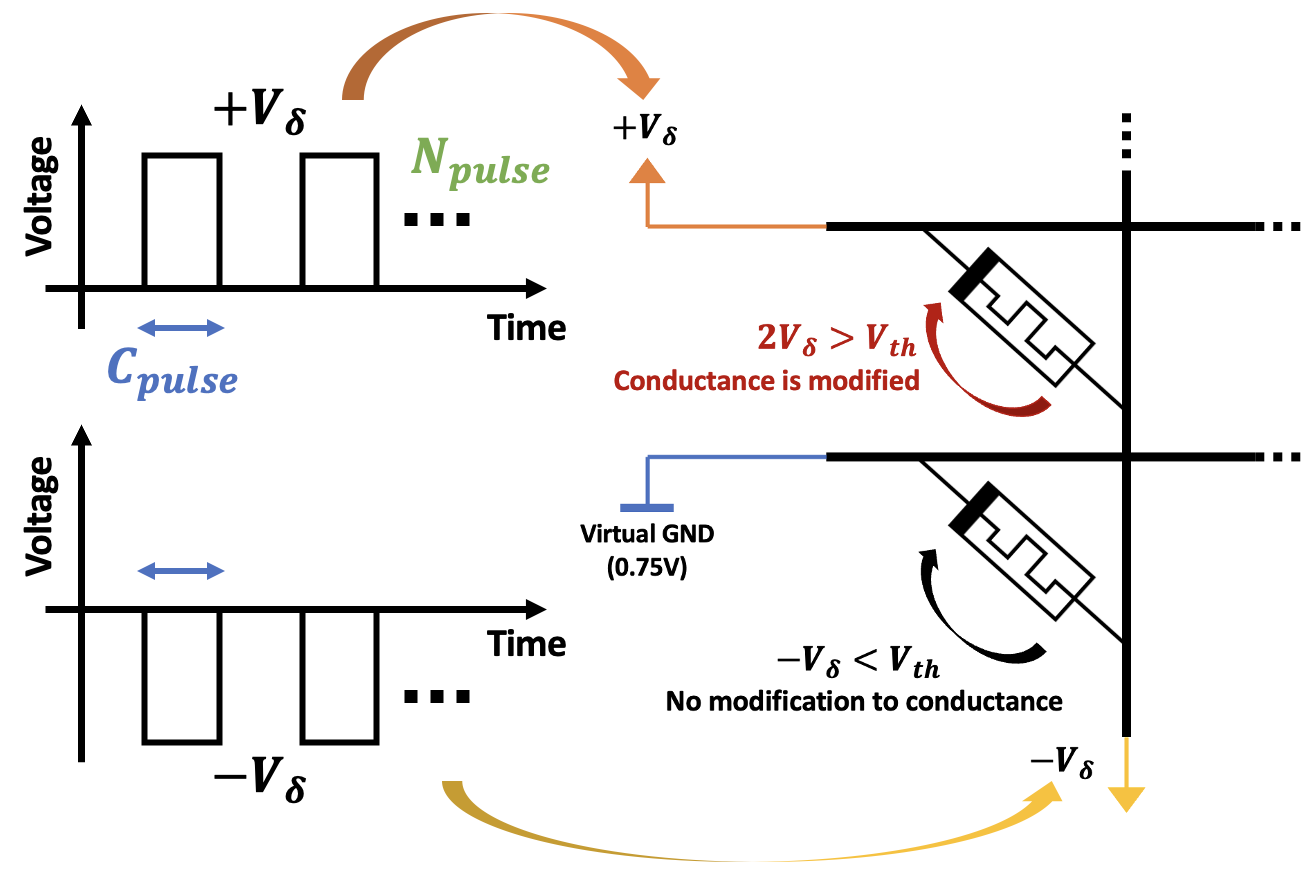}
    \caption{\textit{\textbf{Bipolar memristor conductance control strategy.} }}
    \label{fig:pulses}
\end{figure}

In order to set the conductance values of each memristor in the crossbar during model training, bipolar voltage pulses with amplitudes larger than the memristor threshold voltage $V_{th}$ are utilized. The OpenMENA programming interface enables the setting of the pulse properties (pulse duration command $C_{pulse}$, number of pulses to send $N_{pulse}$ and pulse voltage amplitude $V_{\delta}$) via python scripts. A bipolar pulsing strategy (see Fig. \ref{fig:pulses}) applied at both the input and output of the crossbar is used in order to minimize the cross-modification of neighboring memristors when modifying a target memristor with crossbar coordinate $(x,y)$ \cite{8350906}. This strategy enables the application of a voltage larger than $V_{th}$ across a target memristor $(x,y)$ while keeping the voltage across neighboring memristors below $V_{th}$. 

\section{Training and Weight Setting Algorithms}
\label{traininf}

\subsection{Software model training via constrained optimization}
\label{softwarebased}
In OpenMENA, all weights are strictly positive since memristor conductances can only assume positive values. Therefore, off-chip training must enforce $\Phi_i \geq 0, \forall i$. To incorporate this constraint, we adopt a \textit{Sequential Least Squares Quadratic Programming} (SLSQP) optimization scheme for learning the model tensor $\Phi$ given a loss function to minimize. SLSQP solves nonlinear constrained problems through iterative quadratic programming (QP) steps, with positivity constraints enforced via the Lagrangian framework and corresponding \textit{Karush–Kuhn–Tucker} (KKT) conditions, approached iteratively at each QP step.

To illustrate this, let us consider the use of OpenMENA for performing classification via Softmax logistic regression. In this context, the model can be written as:
\begin{equation}
    \Bar{p}_{out} \xleftarrow{} \text{SoftMax}(\Phi \hspace{3pt} \Bar{x}_{in} + \Bar{b})
    \label{modelillustr}
\end{equation}
where $\Phi$ contains the weights that need to be written as memristor conductance values into OpenMENA, $\Bar{b}$ is a bias vector that will be later fine-tuned using OpenMENA in a chip-in-the-loop setting (see Section \ref{chipinloop}), $\Bar{p}_{out}$ is the class output probability vector and $\Bar{x}_{in}$ is the input vector to be cassified. To learn a strictly positive set of parameters $\Phi$ that also minimizes the loss function $\mathcal{L}$ (e.g., cross-entropy) associated to the problem:
\begin{equation}
    \mathcal{L} = \sum_{i=1}^{N_{data}}\text{CrossEntropy}(\Bar{p}_{out,i}, \Bar{y}_i),
\end{equation}
where $N_{data}$ is the number of training samples and $\Bar{y}_i$ is a one-hot label vector associated to data sample $i$, we make use from the \texttt{minimize} command of the \texttt{sklearn} library which supports SLSQP optimization. Doing so, we obtain a set of weights $\Phi$ that need to be programmed into the memristor crossbar of the OpenMENA board. 

\subsection{Writing weights to the OpenMENA board}
\label{writingweight}
To efficiently write the learned weights $\Phi$ into the memristor crossbar, we propose a VIPI control scheme using a PI regulation loop to finely tune each memristor’s conductance. The VIPI controller iteratively applies bipolar voltage pulses with varying duration $C_{pulse}$ to increase or decrease conductance depending on pulse polarity. Crucially, the pulse voltage amplitude $V_{\delta}$ is incremented after every $n_c$ iterations while the error between the current conductance $\phi$ and target $\phi^*$ exceeds a tolerance $\epsilon$. This voltage-incremental approach proved to be of crucial importance for addressing the variability of memristor threshold voltages. Since some memristors require higher voltages to change conductance, our scheme effectively achieves all target conductances while minimizing unintended modifications in the crossbar (see Section \ref{bipolartech}). Algorithm \ref{alg:VIPI} details the VIPI method for tuning the memristor at coordinate $(x,y)$.
\begin{algorithm}
\caption{VIPI control for memristor weight setting}
\label{alg:VIPI}
\begin{algorithmic}[1]
\REQUIRE Target weight value: $\phi^*$; crossbar coordinate $(x,y)$ of the memristor to be modified; $K_p,K_i$: proportional and integral gain coefficients; $\delta$: voltage increment; $n_c$: number of iteration steps between each voltage increment; $\epsilon$: error tolerance between target and readout weight; $N_{iter}$: total number of iterations.
\STATE $E_{acc} \xleftarrow{} 0$
\STATE $V_{\delta} \xleftarrow{} 0.08$ // Initial weight writing voltage
\FOR{$i = 1$ to $N_{iter}$}
    \STATE $\Bar{I}\xleftarrow{} \{0,0,0,0,0,0,0,0\}$
    \STATE $\Bar{I}[x]\xleftarrow{} 1$
    \STATE $\Bar{I} \xleftarrow{} V^{max}_{read} \times \Bar{I}$ // rescale input voltage to be $<V_{th}$
    \STATE $\Bar{O} \xleftarrow{} \textbf{Infere\_OpenMENA}(\Bar{I})$
    \STATE $\phi \xleftarrow{} \Bar{O}[y]$
    \STATE $E \xleftarrow{} \phi^* - \phi$ // error between target an current weight
    \IF{$|E| < \epsilon$}
        \STATE Break loop and return
    \ENDIF
    \STATE $E_{acc} \xleftarrow{} E_{acc} + E$ // compute integral term
    \STATE $E_{acc} \xleftarrow{} \max(\min(E_{acc}, E_{max}), -E_{max})$ // limiting
    \STATE $C_{pulse} \xleftarrow{} K_p \times E + K_i 
    \times E_{acc}$ // PI controller
    \STATE $\textbf{Write\_Weight\_OpenMENA}(x,y,C_{pulse},V_{\delta})$ 
    
    \IF{$i$ $\%$ $n_c$ = 0}
        \STATE $V_{\delta} \xleftarrow{} V_{\delta} + \delta$ // increment voltage every $n_c$ steps
    \ENDIF

\ENDFOR
\end{algorithmic}
\end{algorithm}

Hence, after the learning of model parameters $\Phi$ (see Section \ref{softwarebased}), we use Algorithm \ref{alg:VIPI} to set the conductance of each memristor $(x,y)$ in the crossbar to their corresponding target conductance $\phi^* = \Phi(x,y)$.

\subsection{Chip-in-the-loop fine-tuning}
\label{chipinloop}

After writing the weights $\Phi$ into the memristor crossbar, an error $E_{tot} = \sum_{x,y} |\phi(x,y) - \Phi(x,y) |$ will persist between the software-based weights $\Phi$ and the actual memristor conductances $\phi$, due to the non-idealities of the hardware writing process (crosstalk between memristors during weight writing, hysteresis, memristor conductance range variability etc.). Hence, OpenMENA's framework features a chip-in-the-loop fine-tuning step described in Algorithm \ref{alg:finetune}. This fine-tuning step embeds the memristor crossbar during the learning of subsequent model parameters such as the bias vector $\Bar{b}$ in the model used in (\ref{modelillustr}). 
\begin{algorithm}
\caption{Chip-in-the-loop model fine-tuning}
\label{alg:finetune}
\begin{algorithmic}[1]
\REQUIRE $D_{train}=\{(\Bar{x}_{in,i}, \Bar{y}_i), i = 1,...,N_{train}\}$: the training dataset with input vectors $\Bar{x}_{in,i}$ and associated labels $\Bar{y}_i$. $\eta$: learning rate for fine-tuning procedure.
\ENSURE The fine-tuned bias vector $\Bar{b}$
\STATE $\Bar{b} \xleftarrow{} random\_normal(size=8)$
\FOR{$j = 1$ to $N_{step}$}
     \STATE $i\xleftarrow{} random\_choice(N_{train})$ // choose data point
     \STATE $\Bar{I} \xleftarrow{} V^{max}_{read} \times \Bar{x}_{in,i}$ // rescale input $<V_{th}$
    \STATE $\Bar{O} \xleftarrow{} \textbf{Infere\_OpenMENA}(\Bar{I})$
    \STATE $\Bar{a}_i \xleftarrow{} \Bar{O}+\Bar{b}$ // add bias to be learned
    \STATE $\Bar{p}_{out,i} \xleftarrow{} \text{SoftMax}(\Bar{a}_i)$
    \STATE $\mathcal{L} \xleftarrow{} \text{CrossEntropy}(\Bar{p}_{out,i}, \Bar{y}_i)$
    \STATE $\Bar{b} \xleftarrow{} \Bar{b} - \eta \frac{\partial \mathcal{L}}{\partial \Bar{b}}$ // compute gradient and fine-tune bias

\ENDFOR
\RETURN $\Bar{b}$
\end{algorithmic}
\end{algorithm}

Using Algorithm \ref{alg:finetune}, subsequent model parameters (such as biases) are fine-tuned in a way such that the non-idealities during the weight writing process are compensated, preventing a significant degradation of model accuracy. Although Algorithm \ref{alg:finetune} focuses on the fine-tuning of the bias vector in (\ref{modelillustr}), it is important to note that a similar approach can be used for the fine tuning of any subsequent model parameters (e.g., other weight matrices and biases) that perform processing on OpenMENA's output.

\section{Experimental Results}
\label{expres}

\subsection{Digit classification}

In this first experiment, we seek to demonstrate the ability of OpenMENA to classify handwritten digit images. To do so, we makeuse of the \texttt{sklearn} \textit{digits} dataset which features $8\times8$ images of handwritten digits. We restrict the dataset to the binary classification of digits representing "0" and "1". Since the memristor crossbar found in OpenMENA is of dimensions $8\times8$, input vectors that can be fed into the crossbar must be of dimension $8$. Hence, we first flatten all $8\times8$ images into $64$-dimensional vectors and then apply PCA decomposition, keeping only the first 8 principal components $\Bar{x}_{in,i}$ for each data point $i$. We then rescale the obtained vectors $\Bar{x}_{in,i}$ such that their values lie between $0$ and $1$ (corresponding to OpenMENA's valid input range):
\begin{equation}
    \Bar{x}_{in,i} \xleftarrow{} \frac{\Bar{x}_{in,i} - \min_i(\Bar{x}_{in,i})}{\max_i(\Bar{x}_{in,i}) - \min_i(\Bar{x}_{in,i})}, \forall i
\end{equation}

\begin{figure}[htbp]
    \centering
    \includegraphics[scale=0.44]{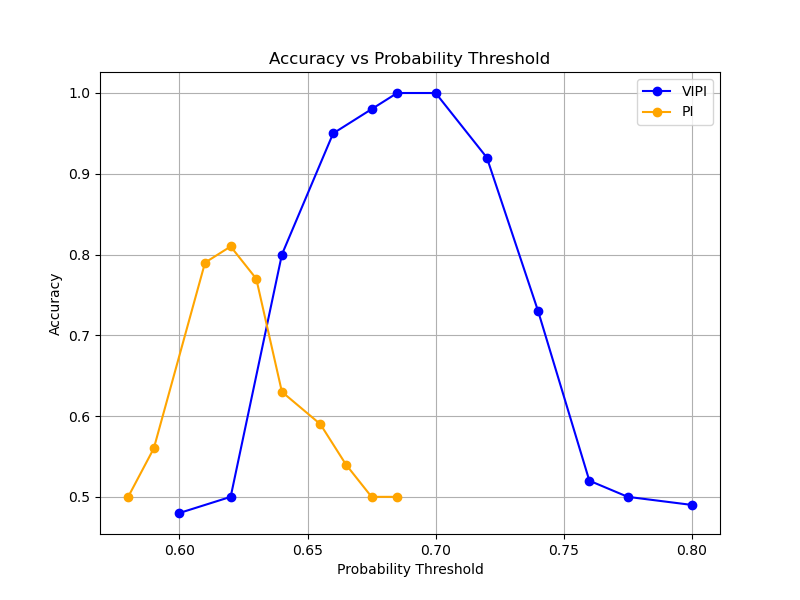}
    \caption{\textit{\textbf{Digit classification test accuracy in function of the model's output decision threshold.} Test accuracy is reported both for our proposed VIPI method and for conventional PI control.}}
    \label{fig:threshs}
\end{figure}

Then, we randomly split the dataset into a $70\%$-$30\%$ train-test split and use the proposed SLSQP-based constrained optimization scheme (see Section \ref{softwarebased}) to learn a set of weights from the training data. Our proposed VIPI scheme (see Section \ref{writingweight}) is then used to write the learned weights into OpenMENA's memristor crossbar, followed by a chip-in-the-loop fine-tuning step for refining the bias value $b$ (see Section \ref{chipinloop}). Finally, we use the held-out test set to characterize the accuracy of the system for different decision thresholds used at the output of the network (see Fig. \ref{fig:threshs}).

Fig. \ref{fig:threshs} clearly demonstrates OpenMENA's ability for signal classification. Furthermore, Fig. \ref{fig:threshs} also shows that our proposed VIPI scheme outperforms the use of previously-proposed PI-based control schemes \cite{e25050774, 10.3389/fnins.2025.1516971} for setting the conductance of memristors within the crossbar, by achieving $\sim +19\%$ in terms of classification accuracy. This is due to the fact that our proposed VIPI scheme takes into account the variability between the memristor threshold voltages and caters for it by gradually increasing the voltage pulse amplitude during the writing of the conductance values. 

\subsection{Robot obstacle avoidance}

\begin{figure}[htbp]
    \centering
    \includegraphics[scale=0.35]{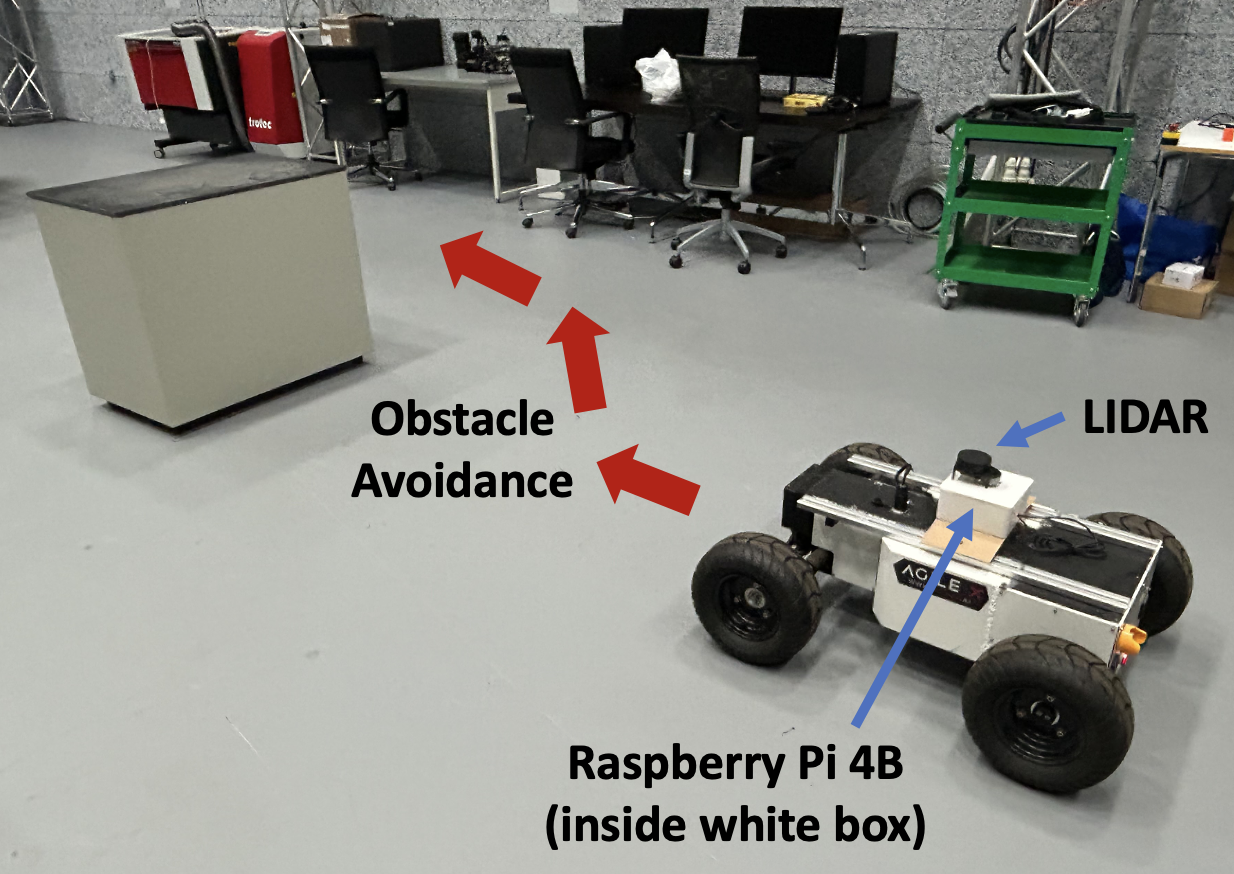}
    \caption{\textit{\textbf{Robot obstacle avoidance task.} A robot car is equipped with a LIDAR connected to as Raspberry Pi 4B which estimates the position of the robot from the LIDAR readings and communicates position and LIDAR data to a remote laptop connected to OpenMENA which sends back control commands to the robot after memristor processing. }}
    \label{fig:robodemo}
\end{figure}

In this second experiment, we investigate how memristor-based processing can be used within a Multi-Layer Perceptron (MLP) model that predicts wheel control commands based on the robot localization data for the task of obstacle avoidance (see Fig. \ref{fig:robodemo}). We first collect a training data sequence consisting of robot position coordinates $x,y$ and yaw angle $\theta$: \([x_t, y_y, \theta_t]\), together with the corresponding motor command labels (forward velocity and steering angle) at each time step \(t\). Then, a 2-layer MLP with 8 neurons per hidden layer using Rectified Linear Unit (ReLU) activations is set up and trained to predict the motor commands based on the positional data. During testing and inference, the first hidden layer of the MLP is implemented through OpenMENA's 8-by-8 memristor crossbar. The second layer is then fine-tuned using the chip-in-the-loop approach of Algorithm \ref{alg:finetune}. 

As input to the model, we first normalize all position and yaw angle data \([x_t, y_t, \theta_t]\) to lie in the range \([0,1]\), corresponding to the subthreshold inference voltage range of $[0,50]$-mV in OpenMENA. The normalized localization vectors are then used as input to the memristor crossbar within the MLP model. 

\begin{figure}
    \centering
    \includegraphics[width=1\linewidth]{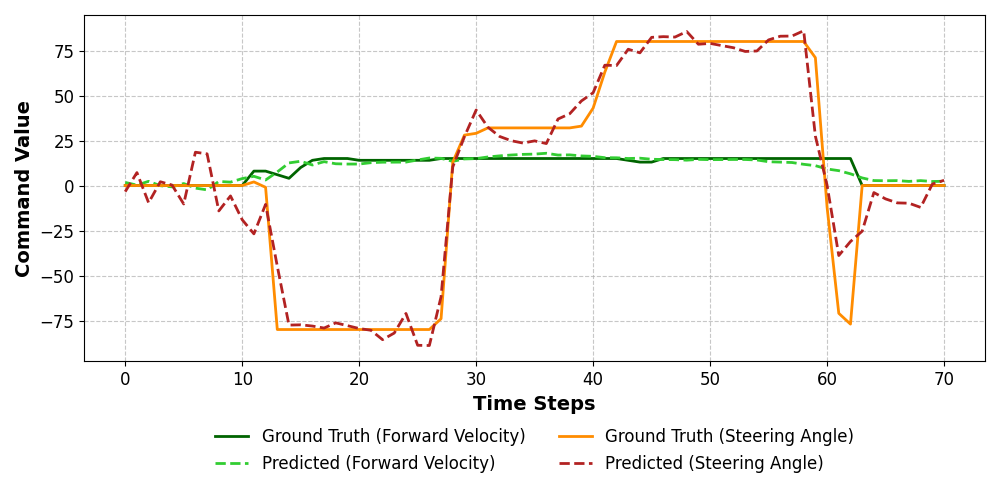}
    \caption{\textbf{\textit{Comparison between ground-truth and predicted robot commands obtained using the memristor–MLP model.} } }
    \label{fig:compa}
\end{figure}

During real-time inference, the robot continuously sends its localization data \([x_t, y_t, \theta_t]\) to a laptop connected to the OpenMENA board via USB. The laptop preprocesses this data by converting it into voltage inputs, which are applied to the memristor crossbar as the first hidden MLP layer. The resulting output vector is read and passed to the second MLP layer (in software), completing the prediction of control commands (forward velocity and steering angle). These predicted commands are transmitted to the robot via WiFi to update its motion. Fig. \ref{fig:compa} shows the predicted velocity and steering angle from our memristor-based MLP, closely matching the ground-truth commands with a low root MSE deviation of $8.7$. 
Finally, a video showcasing our memristor-controlled robot setup is provided in the Supplementary Material. 



\section{Conclusion}
\label{conc}

This paper presented OpenMENA, a fully open-source memristor interfacing and compute board for exploring analog neuromorphic edge AI design. OpenMENA supports 8×8 memristor crossbar arrays and includes a companion Python library for easy programming and control. We also introduced an SLSQP-based constrained training approach and  VIPI method for writing learned weight matrices into the crossbar. Experiments demonstrated OpenMENA’s use in digit classification and robot obstacle avoidance. We hope OpenMENA will democratize memristor-based AI research and inspire advances in continual edge learning and synaptic plasticity–based local learning.

 \bibliographystyle{IEEEtran}
\bibliography{refs}

\end{document}